\documentclass[twocolumn,prl,aps,showpacs,floatfix,]{revtex4}

\usepackage{graphicx}

\begin{document}

\bibliographystyle{apsrev}

\title{Silicene on metallic quantum wells -- an efficient way of tuning
silicene-substrate interaction}

\date{\today}

\author{A. Podsiad\l y-Paszkowska}
\affiliation{Institute of Physics, Maria Curie-Sk\l odowska University, 
             pl. M. Curie-Sk\l odowskiej 1, 20-031 Lublin, Poland}

\author{M. Krawiec}
  \email{mariusz.krawiec@umcs.pl}
\affiliation{Institute of Physics, Maria Curie-Sk\l odowska University, 
             pl. M. Curie-Sk\l odowskiej 1, 20-031 Lublin, Poland}

\begin{abstract}
We propose a powerful method of controlling interaction between silicene and
a substrate utilizing quantum size effect, which allows to grow silicene with 
tailored electronic properties. As an example we consider silicene on ultrathin 
Pb(111) layers, and demonstrate how the properties of silicene, including the 
binding energy, and the Dirac bands, can easily be tuned by quantum well states 
of the substrate. We also discover a novel mechanism of protecting the Dirac 
electrons from the influence of the substrate. This is associated with special 
arrangement of a part of Si atoms in silicene. These findings emphasize the 
essential role of interfacial coupling and open new routes to create 
silicene-like two-dimensional structures with controlled electronic properties.
\end{abstract}


\maketitle

Silicene, the two-dimensional (2D) allotrope of silicon, has attracted 
considerable attention due to its similar to graphene exceptional electronic 
properties \cite{Takeda1994,Cahangirov2009}. However, silicene and graphene 
differ in the atomic structure -- in both of them atoms are ordered in the 
honeycomb lattice, but due to sp$^2$ hybridization, graphene is entirely flat,
while mixing of sp$^2$ and sp$^3$ hybridization in silicene leads to 
low-buckled structure. Due to this buckled structure a band gap can be easily 
generated and controlled by electric field \cite{Ni2012,Drummond2012,Tsai2013}. 
Properties of silicene can also be tuned by chemical functionalization 
\cite{Osborn2011,Lin2012,Zheng2012}.

Since silicon tends to hybridize in sp$^3$ form there is no graphite-like
layered material composed of silicon atoms, thereby epitaxy is the only way to 
obtain the silicene layer, as it was successfully demonstrated on a few 
different substrates 
\cite{Vogt2012,Fleurence2012,Meng2013,Morshita2013,Aizawa2014}. However, 
presence of the substrate may and usually does alter the properties of the
free-standing silicene. Indeed, despite extensive studies of silicene on Ag(111)
surface 
\cite{Vogt2012,Jamgotchian2012,Chiappe2012,Majzik2013,Chen2012,Wang2013,Guo2013,
Lin2013,Feng2013} problem of band structure (presence or absence of linear
bands) does not yet have a precise answer. There are firm arguments that 
binding of silicene to Ag surface is so strong that it destroys not only the 
Dirac cone but also 2D character of silicene. However, a recent paper of Y. 
Feng et al. \cite{Feng2015} indicates that Dirac cones survive and appear in K 
points of silicene 3$\times$3 Brillouin zone. These studies clearly point to 
the importance of the silicene-substrate interaction. 

Thus searching of alternative substrates for the silicene growth is a still 
debated issue. Semi-conducting surfaces seem to be particularly attractive
in view of their weak interaction with silicene layer, which is beneficial for 
preserving the linear nature of the band structure 
\cite{Guo2013,Liu2013,Bhattacharya2013,Zhu2014,Gao2014,Kokott2014}. However,
such weak interaction will likely result in clustering of silicon atoms, and 
may prohibit formation of 2D Si layer at all. In the case of metal substrates, 
the silicene-substrate interaction is too strong to maintain the properties of 
the free-standing silicene, and leads to substantial modifications or to 
destruction of the Dirac-fermion spectrum \cite{Pflugradt2014}. It seems that 
the best choice would be the substrate featuring moderate interaction with 
silicene. Recently, we proposed to use lead as a substrate, a material which 
seems to fulfill the above interaction requirement \cite{Podsiadly2015}. Indeed, 
in this case, the binding energy falls between energies characteristic of 
silicene on a typical semiconductor and on a metal substrate. As a result, the 
Dirac $\pi$ bands are only slightly modified with the main contribution of the 
silicene 3p$_z$ orbitals. 

Searching for new templates is one of the most natural directions in the field
of silicene formation. However, it is rather laborious, often complicated, and
not much effective. In the present work, we propose an alternative approach, 
utilizing quantum phenomena rather than new materials, to form silicene, and to 
control its properties. The idea is to use as a substrate metallic quantum 
wells (QW), i.e. ultrathin metal layers, in which a quantum size effect (QSE) 
takes place. It is well known that the QSE influences the properties of thin 
metal films and makes the quantities related to atomic structure, energetics 
and electronic properties oscillatory functions of a number of monolayers
\cite{Schulte1976,Jalochowski1992,Materzanini2001,Tringides2007,Dil2007,
Slomski2011}. 

The positions of QW states are determined by thickness of the slab, which is 
very attractive in a view of silicene growth. It should be possible to 
manipulate the strength of the silicene-QW interaction, thus to allow to grow 
silicene with tailored electronic properties. As an example we consider 
silicene on ultrathin Pb(111) layers, and demonstrate how the properties of 
silicene, including binding energy, and the Dirac cone, can easily be modified 
by QW states. The proposed idea can be applied to other 2D materials, like
germanene, stanene, etc., on different QSE substrates. Furthermore, we also 
discover novel mechanism of protecting the Dirac electrons from influence of 
the substrate, which is associated with the presence of Si atoms sticking out 
of silicene layer. These findings emphasize the essential role of interfacial 
coupling and open new routes to create 2D structures with controlled electronic 
properties.

The calculations were performed within density functional theory using SIESTA 
code \cite{Ordejon1996,Sanchez1997,Artacho1999,Soler2002,Artacho2008}. The 
generalized gradient approximation in Perdew-Burke-Ernzerhof form 
\cite{Perdew1996} was utilized as the exchange-correlation functional. The 
electron-ion interactions were represented by the Troullier-Martins 
norm-conserving pseudopotential \cite{Troullier1991}. The plane wave cutoff for 
all calculations was set to 200 Ry. The Brilloiun zone was sampled by 
$12 \times 12 \times 1$ k points, according to the Monkhorst-Pack scheme
\cite{Monkhorst1976}.

To avoid additional strain we used superlattices $3\times3$ of Pb(111) surface 
and $\sqrt7\times\sqrt7$ of silicene with approximately 3\% of lattice 
mismatch. The slab contains different number of Pb monolayers (from 2 to 8) and 
a silicene layer on the top. Only atoms from the lowest layer were fixed in 
their bulk positions and the rest of the system was fully relaxed until all 
the forces acting on atoms were lower than 0.01 eV/\AA\ . The initial geometry 
of each QW system was the corresponding relaxed Pb slab with the bulk in-plane 
lattice constant and the free-standing silicene layer. 

The STM simulations have been performed within the Tersoff-Hamann approach in 
the constant-current mode \cite{Tersoff1983,Tersoff1985}. The analysis of 
charge transfer between different atoms has been done according to Bader
\cite{Bader1990}.

\begin{figure}[h]
\includegraphics[width=\linewidth]{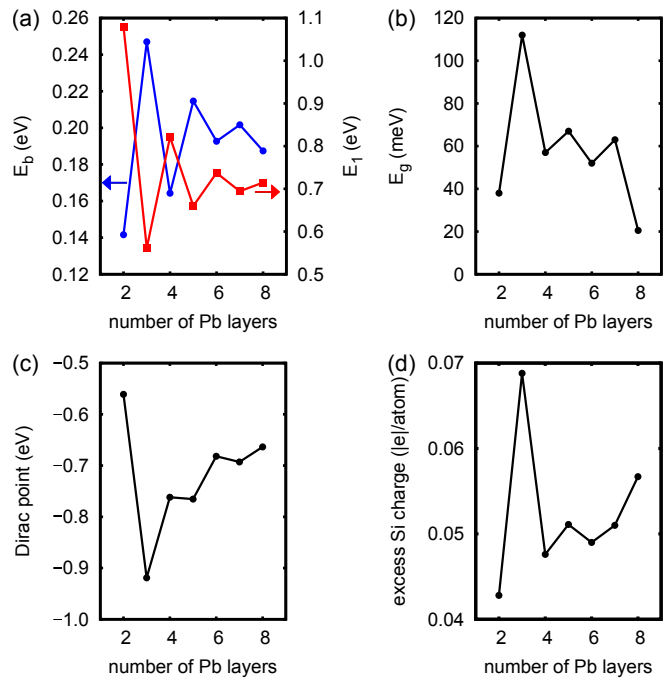}
\caption{\label{Fig1} (Color online) (a) The binding energy of silicene to Pb 
         films (blue line), and of top Pb layer to rest of the slab (red line) 
	 as a function of different number of Pb layers. Evolution of energy 
	 gap at the Dirac point (b), position of the Dirac point (c), and 
	 charge transfer between silicene and Pb substrate (d) with Pb QW 
	 thickness.}
\end{figure}

Figure \ref{Fig1} (a) (blue line) shows the binding energy of silicene to 
Pb(111) film composed of a different number of layers.
Clear oscillations of $E_b$ as a function of Pb thickness are visible. This is
the result of different positions of quantized energy levels of Pb film due to 
the quantum size effect (see Fig. 1 in Ref. \cite{support}). Obviously, these 
quantum well states are also responsible for binding energy $E_1$ of the top Pb 
layer to rest of Pb slab [see red line in Fig. \ref{Fig1} (a)]. Thus both, 
$E_b$ and $E_1$, show oscillatory behavior as a function of number of Pb 
layers, but they oscillate in antiphase. It means that if the top Pb layer is 
bound strongly (weakly) to rest of the slab, it binds silicene weakly 
(strongly). This is an example of common wisdom in chemistry on the bonding of 
atoms, i.e. if a given atom is strongly involved in bonding with one of its 
neighbors, at the same time it weakens its bonds with the other neighbor 
\cite{Ossowski2015}. Such behavior of the binding energy can be exploited in 
the growth of silicene layer. Changing the thickness of a slab, thus the 
binding energy, we modify the growing conditions for silicene. Thus we are 
equipped with a new tool to control the formation of silicene. The same 
scenario should be realized in other 2D systems on various thin metal layers, 
provided van der Waals interactions are not much important, and substrates 
exhibit QSE. Obviously, this is unlikely to work for graphene.

Surprisingly, the binding energy oscillations are not reflected in substantial
changes of the silicene atomic structure. The silicene layer is buckled and 
consists of Si atoms in their top and bottom positions \cite{Podsiadly2015}. 
The arrangement of Si atoms is almost the same, independent of QW thickness. 
The influence of QSE leads only to different mean separations of silicene and 
Pb surface, with large (small) distance for weak (strong) silicene-substrate 
interaction, and to small variations of the buckling, only to 0.06 \AA. The 
energy ordering of different silicene reconstructions, predicted in Ref. 
~\citenum{Podsiadly2015}, is not sensitive to QSE either. The lack of 
structural modifications is supported by recent experiment on silicene on 
Ag(111) films, albeit containing more atomic layers \cite{Sone2014}.

\begin{figure*}[htbp]
\includegraphics[width=\linewidth]{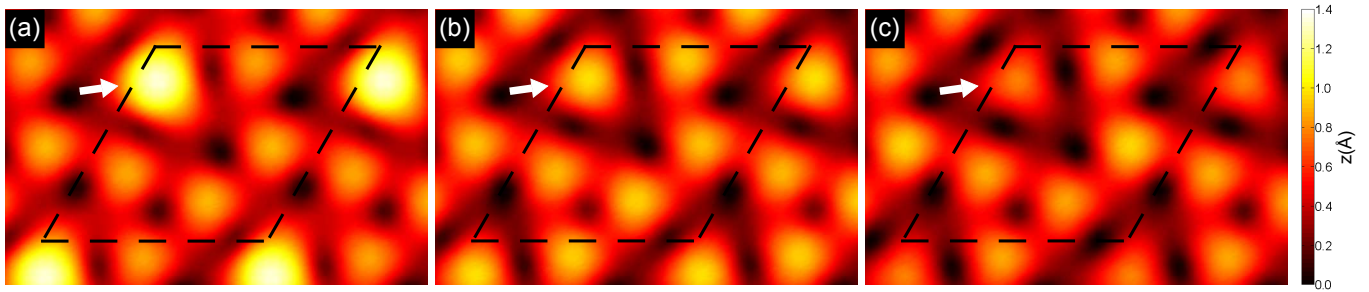}
\caption{\label{Fig2} (Color online) (a) Constant current STM topography of 
         silicene on 3 ML thick Pb film, calculated at the bias voltage -1 V. 
	 (b) and (c) corresponding images for 4 and 5 ML. Arrows point to the 
	 bright protrusions, most strongly affected by QSE.}
\end{figure*}

\begin{figure*}[htbp]
\includegraphics[width=\linewidth]{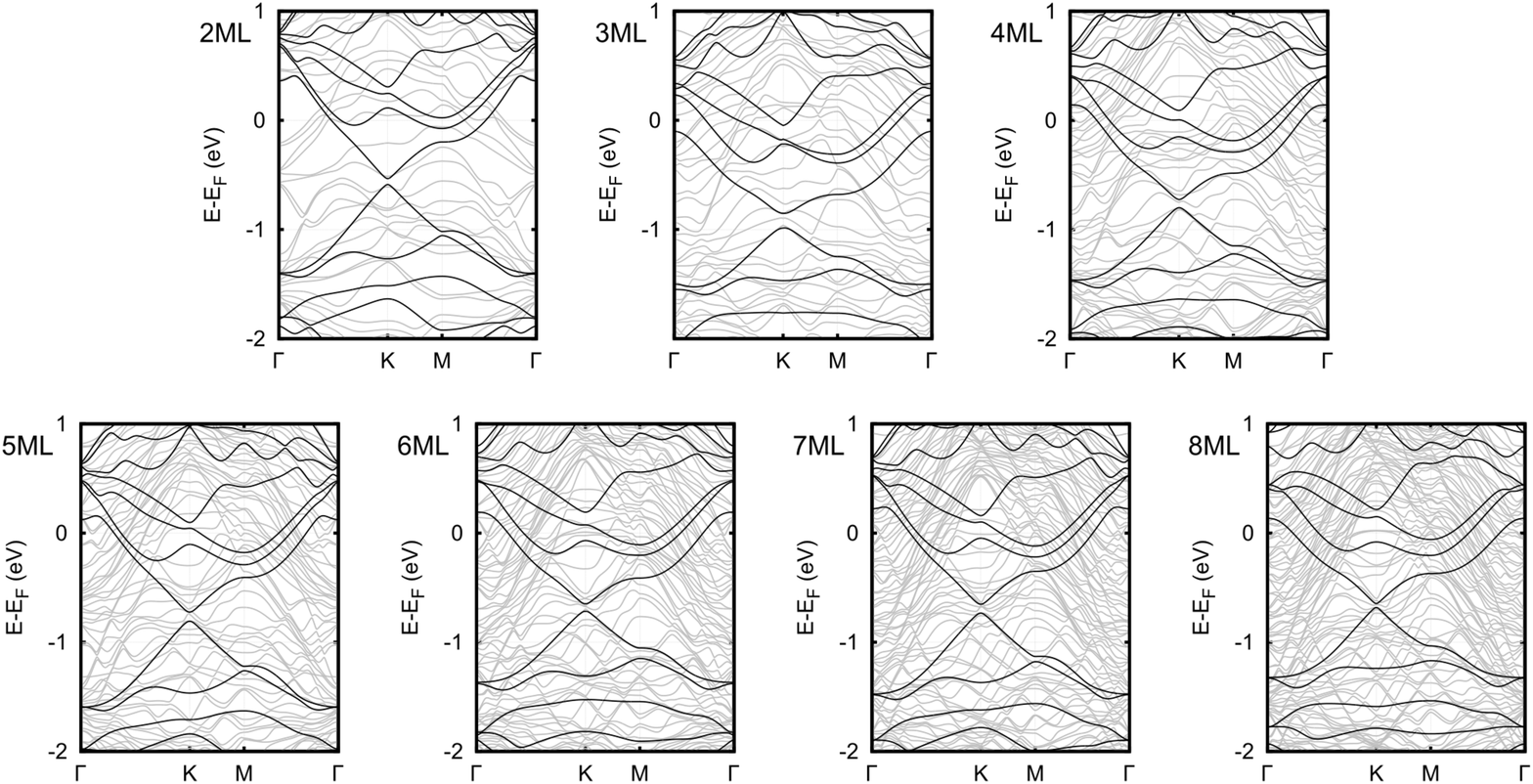}
\caption{\label{Fig3} Evolution of band structure silicene on Pb 
         layers as a function of Pb thickness. Black solid lines show the 
	 unsupported silicene band structure, which has been calculated by 
	 fixing Si atoms positions as obtained in full slab calculations and 
	 removing the substrate.}
\end{figure*}

Scanning tunneling microscopy (STM) topography images calculated for silicene 
on quantum wells of different thickness show, however, noticeable changes, as 
one can deduce from experiments on bare Pb layers \cite{Eom2006}. These changes 
are most pronounced for Si atoms in the top position, as it is visible in Fig. 
\ref{Fig2} for marked bright protrusion.
This suggests that electronic effects play an important role in the system. 
Namely, the QW states of the substrate provoke changes in the electronic 
structure of silicene, thus influence the STM topography. Indeed, this is 
clearly visible in Fig. \ref{Fig3}, where evolution of the energy 
bands as a function of Pb thickness is shown. 
In all cases bands with linear 
dispersion can be detected, albeit modified. Due to the presence of a 
substrate, the inversion symmetry of silicene lattice is broken, which results 
in the opening of an energy gap at the Dirac point. The magnitude of this gap, 
is related to QSE via the binding energy and the mean silicene-Pb surface 
separation, and oscillates as a function of QW thickness [see Fig. \ref{Fig1} 
(b)]. 

Similarly, position of the Dirac point [Fig. \ref{Fig1} (c)], defined as energy 
in the middle of the Dirac cone gap, can be shifted by approximately 0.4 eV by
varying QW thickness. This effect is accompanied by charge transfer between 
both subsystems, as it is evidenced in Fig. \ref{Fig1} (d). However, the 
situation is more complicated, and additional mechanism must be present, since 
this charge transfer is too low to shift the Dirac point so deep in the 
valence bands. Indeed, not all Si atoms receive negative charge, as the Bader 
analysis indicates. Namely, Si atoms in the lower layer, which contribute 
mainly to the Dirac cone, are doped by electrons from the top Si layer, which 
further shifts the Dirac cone down. Thus the electron charge is transferred 
from both Pb and top Si atoms to the bottom Si atoms. Moreover, there is strong 
hybridization of electron states, mainly derived from p-orbitals of Si 
($\sigma$ and $\pi$ bands of silicene) and Pb atoms, in vicinity of the 
$\Gamma$ point. This can be observed while comparing unsupported silicene bands 
with the band structure of the full slab near the $\Gamma$ point. Clearly, 
strong modifications of the unsupported silicene bands are visible. Note that 
the strain imposed on silicene layer is too weak to shift the Dirac cone
\cite{Wang2013b}.

\begin{figure*}[t]
\includegraphics[width=0.9\linewidth]{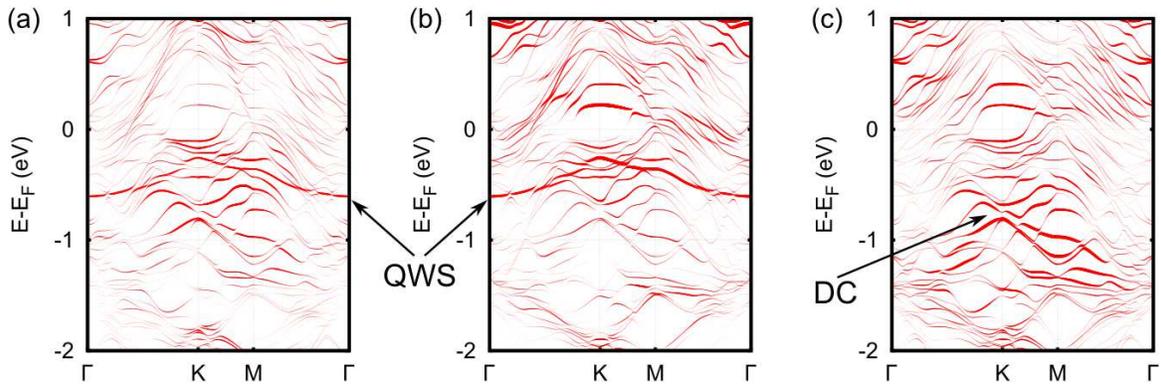}
\caption{\label{Fig4} (Color online) Projected band structure of silicene on 4 
         ML thick Pb film. (a) Contribution of Pb 6p$_z$, (b) top-Si 3p$_z$, 
	 and (c) bottom-Si 3p$_z$ states to the electronic bands. QWS and DC
	 stand for the quantum well state and the Dirac cone, respectively.}
\end{figure*}

The charge transfer between top and bottom Si atoms not only shifts the Dirac 
cone, but is an intrinsic factor of a novel effect, which we call 
{\it self-protection of silicene}. The important role in the mechanism of
self-protection falls to the top Si atoms in supported silicene layer. Namely, 
the $\pi$ electrons of silicene try to avoid the destructive influence of a 
substrate as much as they can for the price of strong interaction between 
top Si atoms with Pb film. Such behavior is triggered by the QSE, as the 
electron wave function forms a standing wave limited by bottom of the slab and 
the top-most objects (top Si atoms). Thus the role of the top Si atoms is to
decouple the Dirac electrons in silicene from other parts of the system.
Consequently, we can think of silicene on a substrate as an electron doped 
quasi-free-standing layer with top Si atoms acting as substitutional 
impurities. This scenario is supported by projected band structure, shown in 
Fig. \ref{Fig4}. 
Clearly, the main contribution to the linear bands comes from the bottom Si 
atoms, with a weak impact of Pb 6p$_z$ and even weaker of top Si 3p$_z$ states. 
On the other hand, there is strong interaction between Pb and top Si p$_z$ 
orbitals, forming QW states. This is a hallmark of the self-protection
mechanism. Note that there is no trace of this QW state in the electronic 
structure of the bottom Si atoms. Thus the Dirac cone in silicene on QW 
survives and the $\pi$ electrons of silicene try to be decoupled from the rest 
of the system owing to strong interaction between Pb and top Si atoms. We think
that the self-protection scenario is realized in other supported 2D systems, 
like silicene on Ag(111), where the atomic structure of silicene also features 
a few Si atoms in the unit cell sticking out of the Si layer. However in this 
case the self-protection mechanism may be too weak to compete with the very 
strong silicene-substrate interaction. 

\begin{figure*}[t]
\includegraphics[width=0.9\linewidth]{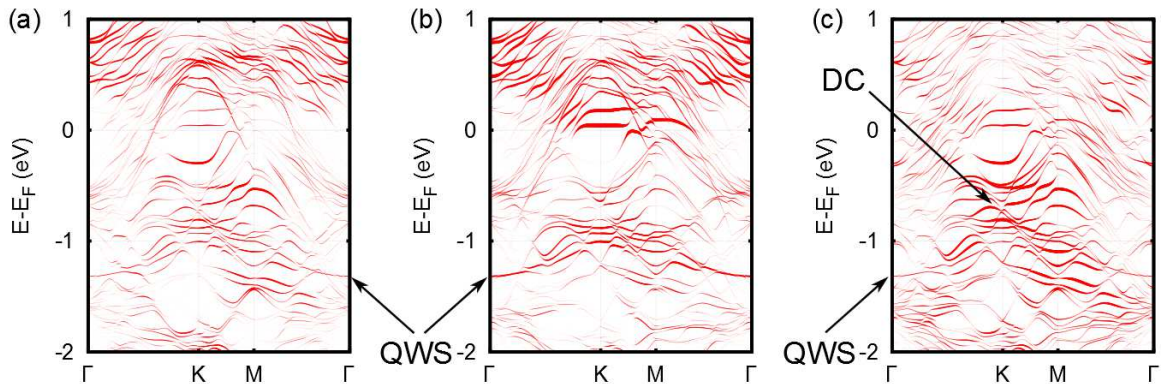}
\caption{\label{Fig5} (Color online) The same as in Fig. \ref{Fig4} for 
         silicene on 5 ML thick Pb film.}
\end{figure*}

The decoupling of $\pi$-electrons depends on relative position of the 
Dirac point and QW state, and can be slightly suppressed if both coincide, as 
it is seen in Fig. \ref{Fig5}. 
This may occasionally happen for certain thicknesses, and rather for thicker 
films, where the separation between QW states is smaller.

Finally, we would like to discuss the experimental realization of the proposed
method of growing and controlling 2D silicene-like systems. The strongest 
effect should be observed for thin films. However, the QWs also require 
substrates. Therefore, it is important to chose the substrate on which the film 
can be grown in a layer by layer mode. Furthermore, present calculations have 
been performed for unsupported QWs. The underlying substrate on which thin 
films could be grown will certainly influence the QW states to some extent. 
Therefore, one can choose a substrate with minor impact on QW states like, for 
example, single-crystalline epitaxial graphite \cite{Dil2007}. On the other 
hand, the supported quantum wells may even help to control the properties of 
silicene, as in the case of Pb films on Si(111) surface, where the QW states 
become less dispersive, and remain flat in a large part of the Brillouin zone 
\cite{Slomski2011}. Thus the silicene properties should be more controllable. 
Finally, we would like to stress that the proposed idea can be utilized in
other 2D materials on various substrates exhibiting QSE.


In conclusion, we proposed a powerful method of controlling the 
silicene-substrate interaction owing to the quantum size effect. We have 
demonstrated that the electronic properties of silicene on metallic quantum
wells, like binding energy, position of the Dirac cone and magnitude of the 
energy gap at the Dirac point, can easily be tuned by quantum well states of 
the substrate. This idea can be utilized in other 2D silicene-like materials 
growing on various thin films. We also discovered the effect of self-protection 
of silicene, in which the top Si atoms play a crucial role, taking on the 
interaction with the substrate, and decoupling Dirac electrons from the rest of 
the system. These findings emphasize the essential role of interfacial coupling 
and open new routes to create two-dimesnional structures with tailored 
electronic properties.

This work has been supported by the National Science Centre under Grant no. 
2014/15/B/ST5/04244.


\end{document}